\documentclass[10pt,conference,letterpaper]{IEEEtran}
\IEEEoverridecommandlockouts
\usepackage[letterpaper, top=1.92cm, bottom=2.9cm, left=0.63in, right=0.621in]{geometry}
\makeatletter
\def\ps@headings{%
\def\@oddhead{\mbox{}\scriptsize\rightmark \hfil \thepage}
\def\@evenhead{\scriptsize\thepage \hfil \leftmark\mbox{}}
\def\@oddfoot{}%
\def\@evenfoot{}}
\makeatother

\usepackage{subfigure}
\usepackage{colortbl}
\usepackage{bm}
\usepackage{cite}
\usepackage{exscale}
\usepackage{relsize}
\usepackage[utf8]{inputenc}

\usepackage{graphicx}  
\usepackage{url}       

\usepackage{amsthm,amsmath,amssymb}   
\usepackage{mathrsfs}
\usepackage{cite}

\usepackage{amsfonts,amssymb}

\usepackage{stfloats}

\usepackage{CJK}
\usepackage{indentfirst}
\usepackage{cases}
\usepackage{algorithm}
\usepackage{multirow}
\usepackage{algorithmic}
\usepackage{stfloats}
\usepackage{colortbl,booktabs}
\usepackage{color}
\usepackage{epstopdf}
\usepackage[absolute,overlay]{textpos}
\usepackage{mdframed}

\makeatletter

\newcommand{\Rmnum}[1]{\expandafter\@slowromancap\romannumeral #1@}
\makeatother

\newcommand{\ls}[1]
    {\dimen0=\fontdimen6\the\font
     \lineskip=#1\dimen0
     \advance\lineskip.5\fontdimen5\the\font
     \advance\lineskip-\dimen0
     \lineskiplimit=1.0\lineskip
     \baselineskip=\lineskip
     \advance\baselineskip\dimen0
     \normallineskip\lineskip
     \normallineskiplimit\lineskiplimit
     \normalbaselineskip\baselineskip
     \ignorespaces
    }

\begin{document}

\newcommand{\copyrightstatement}{
    \begin{textblock*}{\textwidth}(1.5cm, 26cm)
         \noindent
         \footnotesize
         \begin{mdframed}
         \copyright 2025 IEEE. Personal use of this material is permitted. Permission from IEEE must be obtained for all other uses, in any current or future media, including reprinting/republishing this material for advertising or promotional purposes, creating new collective works, for resale or redistribution to servers or lists, or reuse of any copyrighted component of this work in other works. DOI: 10.1109/WCNC61545.2025.10978314.
         \end{mdframed}
    \end{textblock*}
}
\copyrightstatement
\setlength{\textwidth}{0.98\textwidth}

\title{
RSMA Assisted ISAC With Hybrid Beamforming 
}

\vspace{-0pt}
\author{Zhuohui Yao\IEEEauthorrefmark{2}, Wenchi Cheng\IEEEauthorrefmark{2}, Liping Liang\IEEEauthorrefmark{2}, Tao Zhang\IEEEauthorrefmark{2}, and Jun Gong\IEEEauthorrefmark{2}\\

\vspace{-0pt}

\IEEEauthorblockA{$^{\dagger}$State Key Laboratory of Integrated Services Networks, Xidian University, Xi'an, China\\
E-mails: \{\emph{yaozhuohui@xidian.edu.cn}, \emph{wccheng@xidian.edu.cn}, \emph{liangliping@xidian.edu.cn}, and \emph{zhangtao02@xidian.edu.cn}\}}

\vspace{-5pt}

	\thanks{\ls{0.5}This work was supported in part by the National Natural Science Foundation of China (Nos. 62341132 and 62201427) and the Postdoctoral Fellowship Programs of CPSF (Nos. 2023TQ0256 and GZC20232056).}
\vspace{-1cm}
}
\maketitle

\begin{abstract}
The harsh environment and scarce resources post-disaster drive the equipment to be miniaturized and portable. Based on this, integrated sensing and communication (ISAC) systems play a significant role in providing emergency wireless networks. In order to reduce the hardware cost, a hybrid beamforming (HBF) assisted millimeter-wave (mmWave) ISAC system, which exploits the limited number of radio frequency (RF) chains, is considered in this paper.  However, the HBF structure reduces the spatial degrees of freedom, thus leading to increased interference among communication users and radar sensing. To solve this problem, a rate-splitting multiple access (RSMA) strategy is adopted to enhance the emergency mmWave-ISAC system. We formulate the weighted sum rate (WSR) maximization objective by jointly designing common rate allocation and HBF.
Then, we propose the penalty dual decomposition (PDD) coupled with the weighted mean squared error (WMMSE) method to solve this high-dimensional non-convex problem. Numerical results demonstrate the effectiveness of the proposed algorithm and show that the RSMA-ISAC scheme outperforms other benchmark schemes.

\end{abstract}

\vspace{-1pt}


\vspace{-0.5cm}
\section{Introduction}
\IEEEPARstart{T}{he} rapid advancement of urbanization coupled with information construction, has inadvertently heightened the susceptibility of urban lifelines to disruptions in the wake of catastrophic events. After natural disasters, the fundamental infrastructure is often paralyzed\cite{yaoJSAC}. Various items of emergency rescue equipment are taken with rescuers to the front line to meet the timing of disaster information acquisition and transmission. The harsh environment and scarce resources post-disaster drive the equipment to be integrated and portable\cite{IOTJ2023}.  To further increase the probability of survivals during the critical golden-rescue period\cite{yaoNetwork}, the aforementioned equipment must possess both high-throughput communication and high-accuracy sensing capabilities.
However, the exponential growth of emergency services and the massive number of connections to rescue devices are straining spectrum resources, creating an urgent need for additional spectrum. Based on this, millimeter-wave (mmWave) based integrated sensing and communications (ISAC)\cite{ISACLin}, has been considered one of the new paradigms for emergency wireless networks\cite{ZhangJSC}.

To compensate for the shortages of severe path-loss and rain attenuation in mmWave communications, massive multiple-input multiple-output (MIMO) is employed to generate high-gain directional beams\cite{10279443}\cite{yue2024}. To decrease the hardware costs and power consumption of fully digital large-scale antenna array structure,  the hybrid beamforming (HBF) for ISAC systems was designed to connect the large-scale antenna arrays via phase shifters (PSs) with fewer radio frequency (RF) chains\cite{10582895}. However, the adoption of HBF introduces challenges related to managing interference among communication users and radar sensing. Rate-splitting multiple access (RSMA) is a non-orthogonal transmission strategy based on rate-splitting (RS) precoding at the transmitter and successive interference cancellation (SIC) decoding at the receiver, which provides strong anti-interference capability for multi-antenna wireless networks \cite{MaoYijie}. RSMA reconciles two extreme interference management strategies, spatial division multiple access (SDMA), which treats interference entirely as noise, and non-orthogonal multiple access (NOMA), which decodes interference completely, to achieve higher spectrum and energy efficiencies. Further, the authors of \cite{XuChengcheng} demonstrated the advantages of RSMA in managing interference between communication users and interference between dual functions of ISAC systems.

Motivated by the advantages of mmWave hybrid architecture's low power consumption and robust RSMA interference management, we introduce RSMA in mmWave ISAC systems to address the joint common rate allocation and HBF design issues to fill the research gap in this area. We further improve the performance trade-off region of the ISAC system and reveal the superiority of the RSMA common data streams at the dual functions of communication and sensing.

The remainder of this paper is organized as follows.
Section II gives the system model and problem formulation of RSMA-assisted emergency mmWave ISAC with HBF structure. The specific algorithm for solving the optimization problem is presented in Section III. The numerical results are presented in Section IV. Finally, our conclusions are provided in Section V.


\section{System Model and Problem Formulation}
\begin{figure}[t]
	\centering
	\includegraphics[width=0.45\textwidth]{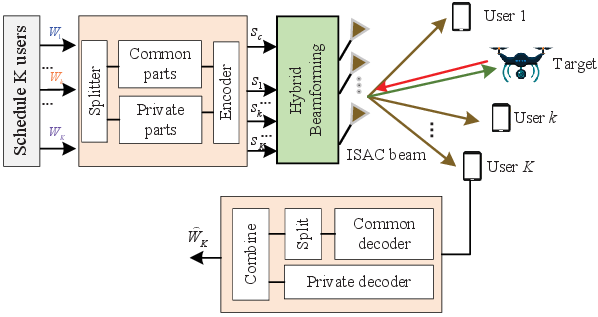}
	\caption{RSMA-assisted emergency mmWave ISAC system.} \label{fig:RSMA-assisted mmWave ISAC system}
\vspace{-0.5cm}
\end{figure}

In this section, we consider an RSMA-assisted emergency mmWave ISAC system as shown in Fig.~\ref{fig:RSMA-assisted mmWave ISAC system}, where one base station (BS) with $N_t$ transmit antennas and $N_r$ receive antennas simultaneously serve $K$ single-antenna downlink user communications while engaging in radar target detection. Let $\mathcal{K}=\{1,2,\cdots, K\}$ represent the set of all user indexes. Specifically, the messages of $K$ users are first split and encoded to form $K+1$ data streams via the RSMA technique.  Without loss of generality, the number of RF chains is ${N}_{\rm RF}$ with the constraint $K+1 \leq {N}_{\rm RF} \leq N_t$.
%

Since the RSMA transmission scheme is adopted, the message $W_k$ of the $k$-th user is first split into two parts: one of which is the common part $W_{c,k}$, and the other is the private part $W_{p,k}$, $\forall k\in\mathcal{K}$. The common data stream $s_{c}$ is obtained by encoding the common message $W_{c}$ formed by combining all user common parts $\{W_{c,1},\cdots,W_{c,K}\}$, while private data streams $\{s_{1},\cdots,s_{K}\}$ are obtained by encoding the private parts $\{W_{p,1},\cdots,W_{p,K}\}$. Define the vector $\boldsymbol{\rm s}=[s_{c}, s_{1}, \cdots ,s_{K}]^{T} \in \mathbb{C}^{(K+1)\times 1}$ as data streams with unit power and satisfy $\mathbb{E}\{\boldsymbol{\rm s}\boldsymbol{\rm s}^{H}\}=\boldsymbol{\rm I}_{K+1}$. Next, the data stream $\boldsymbol{\rm s}$ undergoes hybrid beamforming to obtain a dual-function integrated signal that simultaneously achieves multi-user communication and radar target detection. Accordingly, the transmit ISAC signal, denoted by $\boldsymbol{\rm x}\in \mathbb{C}^{{N_t} \times 1}$, can be expressed as
\vspace{-0.1cm}
\begin{equation}
\begin{aligned}
\boldsymbol{\rm x}\!=\!\boldsymbol{\rm F}_{\rm RF}\boldsymbol{\rm F}_{\rm D}\boldsymbol{\rm s}\!=\!\boldsymbol{\rm F}_{\rm RF}\boldsymbol{\rm f}_{c} s_{c}\!+\!\sum_{i \in \mathcal{K}}\boldsymbol{\rm F}_{\rm RF}\boldsymbol{\rm f}_{{\rm D},i} s_{i},
\end{aligned}
\end{equation}
where $\boldsymbol{\rm F}_{\rm D}=[\boldsymbol {\rm f}_{c}, \boldsymbol {\rm f}_{\rm D,1}, \cdots ,\boldsymbol {\rm f}_{{\rm D},K}]\in \mathbb{C}^{N_{\rm RF} \times (K+1)}$ is the digital beamforming matrix, in which $\boldsymbol {\rm f}_{c}$ is the common data stream precoding vector and $\boldsymbol {\rm f}_{{\rm D},k},k\in\mathcal{K}$ is the precoding vector for the $k$-th user's private data stream. $\boldsymbol{\rm F}_{\rm RF}\in\mathbb{C}^{N_{t} \times N_{\rm RF}}$ is the analog beamforming matrix that is composed of $N_{t}\! \times\! N_{\rm RF}$ phase shifters with a fully connected structure \cite{YuXianghao}. As the phase shifter can only adjust the phase of the signal, all entries in the $\boldsymbol{\rm F}_{\rm R F}$ should satisfy the unit modulus constraints, e.g., $\left|\boldsymbol{\rm F}_{\rm R F}(i, j)\right|=1$.
\vspace{-0.2cm}
\subsection{Communication Model And Metrics}
\vspace{-0.1cm}
Under the above transmit signal model, the received signal, denoted by $y_{k}$, at $k$-th single antenna user can be expressed as
\vspace{-0.1cm}
\begin{equation}
\begin{aligned}
y_{k} =\boldsymbol{\rm h}_{k}^{H} \boldsymbol{\rm F}_{\rm R F} \boldsymbol {\rm f}_{c}s_{c}\!+\!\sum_{i \in \mathcal{K}} \boldsymbol{\rm h}_{k}^{H} \boldsymbol{\rm F}_{\rm R F} \boldsymbol{\rm f}_{{\rm D},i} s_{i}+n_{k},
\end{aligned}
\end{equation}
where ${\boldsymbol{\rm h}}_{k}\in\mathbb{C}^{N_{t} \times 1}$ is the channel vector from the transmitter to the $k$-th user and is assumed to be known by the BS. $n_{k} \sim \mathcal{C N}\left(0,\sigma_{{k}}^{2}\right)$,$k\in\mathcal{K}$, is the additive white Gaussian noise (AWGN) with zero mean and variance $\sigma_{{k}}^{2}$. Without loss of generality, we adopt the widely used Saleh-Valenzuela channel for mmWave communications \cite{Akdeniz}, so $\boldsymbol{\rm h}_{k}$ can be written as $\boldsymbol{\rm h}_{k}=\sqrt{\frac{N_t}{L_k}}\sum_{\ell =1}^{L_k} \beta_{\ell,k}\boldsymbol{\rm a}\left(\theta_{\ell,k}\right)$, where $L_k$ is the number of channel paths for the $k$-th user. $\beta_{1,k}$ and $\beta_{\ell,k},(2\leq\!\ell \!\leq L_k)$ denote the complex gains of the $k$-th user line-of-sight (LoS) component and  non-line-of-sight (NLoS) components, respectively. $\theta_{\ell,k}\in[-\frac{\pi}{2},\frac{\pi}{2}]$ is the angle of departure (AOD) of the $l$-th path, and the array steering vector $\mathbf{a}(\theta)\in\mathbb{C}^{N_{t} \times 1}$ can be given by
$\mathbf{a}(\theta)\!=\!\frac{1}{\sqrt{N_{t}}}\big[1, e^{j \frac{2 \pi}{\lambda} d \sin (\theta)}, \cdots, e^{j\left(N_{t}-1\right) \frac{2 \pi}{\lambda} d \sin (\theta)}\big]^{T}$, where $\lambda$ and $d$ are the carrier wavelength and the antenna spacing, respectively. In this paper, we use a half-wavelength uniform linear array (ULA), i.e., $d=\lambda/2$.

According to \cite{MaoYijie}, the signal-to-interference-plus-noise (SINR) of  decoding common data stream $s_{c}$ and private data stream $s_{k}$ at the $k$-th user, denoted by $\gamma_{c,k}$ and $\gamma_{k}$, respectively, can be given as follows:
\vspace{-0.2cm}
\begin{equation}
\begin{cases}
&{\gamma}_{c,k}=\frac{\left|\mathbf{h}_{k}^{H} \mathbf{F}_{\rm R F} \mathbf{f}_{c}\right|^{2}}{\sum\limits_{i \in \mathcal{K}}\left|\mathbf{h}_{k}^{H} \mathbf{F}_{\rm R F} \mathbf{f}_{{\rm D}, i}\right|^{2} + \sigma_{{k}}^{2}}.\\
&{\gamma}_{k}=\frac{\left|\mathbf{h}_{k}^{H} \mathbf{F}_{\rm R F} \mathbf{f}_{{\rm D}, k}\right|^{2}}{\sum\limits_{i \in \mathcal{K},i\neq k}\left|\mathbf{h}_{k}^{H} \mathbf{F}_{\rm R F} \mathbf{f}_{{\rm D}, i}\right|^{2} + \sigma_{{k}}^{2}}.
\end{cases}
\end{equation}

Based on Eq.(3), the achievable rates of decoding $s_{c}$ and $s_{k}$ for user $k$ are $R_{c,k}=\log_{2} \left(1+\mathrm{\gamma}_{c,k}\right)$ and $R_{k}=\log_{2} \left(1+\mathrm{\gamma}_{k}\right)$, respectively. To ensure that all users can successfully decode the common message, the rate of the common data stream $s_{c}$ meets $ R_{c}={\rm min}\{R_{c,1},\cdots, R_{c, K}\}$ and $\sum_{k \in \mathcal{K}}C_{k}=R_{c}$, where $C_{k}$ is the rate of the common stream $W_{c,k}$ intended for user- $k$. In the RSMA-assisted ISAC system, for the communication function, we select the weighted sum rate (WSR), denoted by $R_{\rm WSR}$, as the performance metric of the system, which can be expressed as $R_{\rm WSR}=\sum_{k \in \mathcal{K}} \alpha_{k}(C_{k}+R_{k})$, where $\alpha_{k}$ denotes the priority factor of the $k$-th user.

\vspace{-0.2cm}
\subsection{Radar Sensing Model And Metrics}
\vspace{-0.1cm}
In ISAC systems, the radar sensing waveform multiplexes the communication signal $\boldsymbol{\rm x}$. We assume that the ISAC system is tasked with detecting a radar target (located at angles $\theta_{0}$) of interest in the presence of interference from $Q$ clutter located at different angles $\theta_{q},q\in\mathcal{Q}=\{1,\cdots, Q\}$.   
 \cite{HassanienAboulnasr}
Thus, the echo signal, denoted by $\boldsymbol{\rm y}_{r}$, can be expressed as
\begin{equation}
\begin{aligned}
\boldsymbol{\rm y}_{r}=&\underbrace{\sqrt{N_{r} N_{t}} \xi_{0} \boldsymbol{\rm a}_{r}\left(\theta_{0}\right) \boldsymbol{\rm a}_{t}^{H}\left(\theta_{0}\right) \boldsymbol{\rm F}_{\rm R F} \boldsymbol{\rm F}_{\rm D} \boldsymbol{\rm s}}_{\text {target }} +\\
&\underbrace{\sqrt{N_{r} N_{t}} \sum_{q \in \mathcal{Q}} \xi_{q} \boldsymbol{\rm a}_{r}\left(\theta_{q}\right) \boldsymbol{\rm a}_{t}^{H}\left(\theta_{q}\right) \boldsymbol{\rm F}_{\rm R F} \boldsymbol{\rm F}_{\rm D} \boldsymbol{\rm s}}_{\text {clutter }}+\boldsymbol{\rm z}_{r},
\end{aligned}
\end{equation}
where $\mathbf{a}_{t}(\theta) \in \mathbb{C}^{N_{{t}} \times 1}$ and $\mathbf{a}_{r}(\theta) \in \mathbb{C}^{N_{{r}} \times 1}$ denote the transmit and receive steering vectors, respectively. $\mathbf{z}_{r} \sim \mathcal{C N}\left(\mathbf{0}_{N_{r}}, \sigma_{{z}}^{2} \mathbf{I}_{N_{r}}\right)$ represents the AWGN with zero mean and variance $\sigma_{{z}}^{2}$. $\xi_{0}$ and $\xi_{q}$ denote the complex fading coefficients of the target and the $q$-th clutter, respectively. 


Then, ISAC BS filters the echo signal $\boldsymbol{\rm y}_{r}$ using the receive beamforming vector $\boldsymbol{\rm v}\in\mathbb{C}^{N_{r} \times 1}$. The filtered signal, denoted by $y_{s}$, can be given by
\vspace{-0.1cm}
\begin{equation}
\begin{aligned}
y_{s}=\xi_{0} \boldsymbol{\rm v}^{H}\boldsymbol{\rm A}\left(\theta_{0}\right)\boldsymbol{\rm x}\!+\!\boldsymbol{\rm v}^{H} \sum_{q \in \mathcal{Q}} \xi_{q} \boldsymbol{\rm A}\left(\theta_{q}\right)\boldsymbol{\rm x}\!+\!\boldsymbol{\rm v}^{H} \boldsymbol{\rm z}_{r},
\end{aligned}
\end{equation}
where $\boldsymbol{\rm A}\left(\theta_{0}\right)=\sqrt{N_{r} N_{t}} \boldsymbol{\rm a}_{r}\left(\theta_{0}\right) \boldsymbol{\rm a}_{t}^{H}\left(\theta_{0}\right)$, $\boldsymbol{\rm A}\left(\theta_{q}\right)=\sqrt{N_{r} N_{t}} \boldsymbol{\rm a}_{r}\left(\theta_{q}\right) \boldsymbol{\rm a}_{t}^{H}\left(\theta_{q}\right)$. Therefore, the filtered output signal-to-clutter-plus-noise ratio (SCNR), denoted by $\gamma_{r}^{out}$, can be expressed as \cite{Tsinos}:
\vspace{-0.1cm}
\begin{equation}
\begin{aligned}
\gamma_{r}^{out }=\frac{\left|\xi_{0} \boldsymbol{\rm v}^{H}\boldsymbol{\rm A}\left(\theta_{0}\right)\boldsymbol{\rm x}\right|^{2}}{\boldsymbol{\rm v}^{H}\left(\boldsymbol{\rm R}_{c}+\sigma_{z}^{2} \boldsymbol{\rm I}_{N_{r}}\right) \boldsymbol{\rm v}},
\end{aligned}\label{eq:gamma_out}
\end{equation}	
where $\boldsymbol{\rm R}_{c}=\sum_{q \in \mathcal{Q}}\left|\xi_{q}\right|^{2} \boldsymbol{\rm A}\left(\theta_{q}\right) \boldsymbol{\rm F}_{\rm RF}\boldsymbol{\rm F}_{\rm D} \boldsymbol{\rm F}_{\rm D}^{H} \boldsymbol{\rm F}_{\rm RF}^{H} \boldsymbol{\rm A}^{H}\left(\theta_{q}\right)$.The receive beamforming problem of solving the output SCNR maximization is well known as the minimum variance distortionless response (MVDR) beamforming problem \cite{Capon}, which has a closed-form solution, denoted by $\boldsymbol{\rm v}^{\star}$, and can be expressed as
$\boldsymbol{\rm v}^{\star}=\frac{\left(\boldsymbol{\rm R}_{c}+\sigma_{z}^{2} \boldsymbol{\rm I}_{N_{r}}\right)^{-1}\boldsymbol{\rm A}\left(\theta_{0}\right)\boldsymbol{\rm x}}{\boldsymbol{\rm x}^{H}\boldsymbol{\rm A}^{H}\left(\theta_{0}\right)\left(\boldsymbol{\rm R}_{c}+\sigma_{z}^{2} \boldsymbol{\rm I}_{N_{r}}\right)^{-1}\boldsymbol{\rm A}\left(\theta_{0}\right)\boldsymbol{\rm x}}$.

By substituting the optimal receive beamforming vector $\boldsymbol{\rm v}^{\star}$ into Eq.~\eqref{eq:gamma_out} and subsequently taking the expectation, we can derive the average output SCNR, denoted as $\overline{\gamma}_{r}^{out }$, which can be expressed as
$\overline{\gamma}_{r}^{out }=\operatorname{tr}\left(\boldsymbol{\rm R}_{0}\left(\boldsymbol{\rm R}_{c}+\sigma_{z}^{2} \boldsymbol{\rm I}_{N_{r}}\right)^{-1}\right)$, where $\boldsymbol{\rm R}_{0}=\left|\xi_{0}\right|^{2}\boldsymbol{\rm A}\left(\theta_{0}\right) \boldsymbol{\rm F}_{\rm RF}\boldsymbol{\rm F}_{\rm D} \boldsymbol{\rm F}_{\rm D}^{H} \boldsymbol{\rm F}_{\rm RF}^{H} \boldsymbol{\rm A}^{H}\left(\theta_{0}\right)$, and $\operatorname{tr}(\cdot)$ denote the trace of the matrix. Based on Cauchy-Schwarz inequality, we scale $\overline{\gamma}_{r}^{out }$ to obtain the lower bound and define it as the input SCNR, denoted by $\overline{\gamma}_{r}^{in}$, given by
\begin{equation}
\begin{aligned}
\overline{\gamma}_{r}^{out}\geq\frac{\operatorname{tr}\left(\boldsymbol{\rm F}_{\rm D}^{H} \boldsymbol{\rm F}_{\rm RF}^{H} \boldsymbol{\rm \Phi} \boldsymbol{\rm F}_{\rm RF}\boldsymbol{\rm F}_{\rm D}\right)} {\operatorname{tr}\left(\boldsymbol{\rm F}_{\rm D}^{H} \boldsymbol{\rm F}_{\rm RF}^{H} \boldsymbol{\rm \Omega} \boldsymbol{\rm F}_{\rm RF}\boldsymbol{\rm F}_{\rm D}\right)+N_{r}\sigma_{z}^{2}}\triangleq\overline{\gamma}_{r}^{in},
\end{aligned}
\end{equation}
where $\boldsymbol{\rm \Phi}=\left|\xi_{0}\right|^{2}\boldsymbol{\rm A}^{H}\left(\theta_{0}\right)\boldsymbol{\rm A}\left(\theta_{0}\right)$ and $\boldsymbol{\rm \Omega}=\sum_{q \in \mathcal{Q}}\left|\xi_{q}\right|^{2} \boldsymbol{\rm A}^{H}\left(\theta_{q}\right) \boldsymbol{\rm A}\left(\theta_{q}\right).$ The input SCNR $\overline{\gamma}_{r}^{in}$ improves the target detection performance even further.

\subsection{Problem Formulation}
In this work, we aim to jointly design the HBF matrix and each common information rate $C_{k},k \in \mathcal{K}$ to maximize the WSR. The optimization problem can be formulated as follows:
\begin{subequations}
	\begin{align}
	\textbf{{{P}}1:}  \max _{\boldsymbol{\rm F}_{\rm R F}, \boldsymbol{\rm F}_{\rm D},\boldsymbol{\rm c}}
	&\sum_{k \in \mathcal{K}} \alpha_{k}(C_{k}+R_{k}) \\
	\text { s.t. } &\frac{\operatorname{tr}\left(\boldsymbol{\rm F}_{\rm D}^{H} \boldsymbol{\rm F}_{\rm RF}^{H} \boldsymbol{\rm \Phi} \boldsymbol{\rm F}_{\rm RF}\boldsymbol{\rm F}_{\rm D}\right)} {\operatorname{tr}\left(\boldsymbol{\rm F}_{\rm D}^{H} \boldsymbol{\rm F}_{\rm RF}^{H} \boldsymbol{\rm \Omega} \boldsymbol{\rm F}_{\rm RF}\boldsymbol{\rm F}_{\rm D}\right)+N_{r}\sigma_{z}^{2}}\geq \gamma_{0}, \label{SCNR} \\
	& \sum_{i \in\mathcal{K}}C_{i}\leq R_{c,k},\forall k \in \mathcal{K}, \label{C}\\
	&\hspace{0.1cm} C_{k}\geq 0,\forall k \in \mathcal{K}, \label{Ck}\\
	&\left\|\boldsymbol{\rm F}_{\rm R F} \boldsymbol{\rm F}_{\rm D}\right\|_{F}^{2} \leq P_{T}, \label{PT}\\
	&\left|\boldsymbol{\rm F}_{\rm R F}(i, j)\right|=1,\forall i,j, \label{units}
	\end{align}
\end{subequations}
where $\left\|\cdot\right\|_{p}$ denote $l_{p}$ norm operation, $\boldsymbol{\rm c}=\big[C_{1},\cdots, C_{K}\big]^{T}$ is a vector that contains the common information rates, $\gamma_{0}$ represents the minimum SCNR threshold to satisfy target detection, and $P_{T}$ is the power budget at the ISAC-BS. Eq. \eqref{C} ensures that each user can successfully decode the common message, and Eq. \eqref{Ck} bounds the nonnegativity of all common rates. Eq. \eqref{units} denotes the unit modulus constraint of the phase shifters.

The nonconvexity of the objective function (OF) complicates finding a global optimum for $\textbf{{{P}}1}$ using conventional optimization algorithms like the block coordinate descent (BCD) and the alternating optimization (AO). 

\section{Proposed WMMSE-PDD Algorithm}
In this section, we propose the WMMSE-PDD algorithm to solve the problem $\textbf{{{P}}1}$. 

\subsection{Problem Reformulation}
According to \cite{XuChengcheng}, the minimizing mean square error (MMSE) of estimating $s_{c}$ and $s_{k}$, denoted by $e_{c,k}^{\rm MMSE}$ and $e_{k}^{\rm MMSE}$ can be given by
\vspace{-0.2cm}
\begin{equation}
\begin{cases}
&\hspace{-0.1cm}e_{c,k}^{\rm MMSE}\!=\!(T_{c, k}\!-\!\left|\boldsymbol{\rm h}_{k}^{H} \boldsymbol{\rm F}_{\rm R F} \boldsymbol {\rm f}_{c}\right|^{2})(T_{c, k})^{-1},\\
&\hspace{-0.1cm}e_{k}^{\rm MMSE}\!=\!(T_{k}\!-\!\left|\boldsymbol{\rm h}_{k}^{H} \boldsymbol{\rm F}_{\rm R F} \boldsymbol {\rm f}_{\rm D,k}\right|^{2})(T_{k})^{-1}.
\end{cases}\label{eq:mmse}
\end{equation}
where $T_{c, k}  =\left|\boldsymbol{\rm h}_{k}^{H} \boldsymbol{\rm F}_{\rm R F} \boldsymbol {\rm f}_{c}\right|^{2}+\sum_{i \in \mathcal{K}}\left|\boldsymbol{\rm h}_{k}^{H} \boldsymbol{\rm F}_{\rm R F} \boldsymbol{\rm f}_{{\rm D},i}\right|^{2}+\sigma_{k}^{2}$ and $T_{k}  =\sum_{i \in \mathcal{K}}\left|\boldsymbol{\rm h}_{k}^{H} \boldsymbol{\rm F}_{\rm R F} \boldsymbol{\rm f}_{{\rm D},i}\right|^{2}+\sigma_{k}^{2}$. Then, we introduce positive weights $u_{c,k}$ and $u_{k}$, the augmented weighted MSEs (AWMSEs) of decoding $s_{c,k}$ and $s_{k}$, denoted by $\eta_{c,k}$ and $\eta_{k}$. By substituting the optimal equalizers and weights into the AWMSEs, we can obtain
$\eta_{c,k}^{\rm MMSE}\triangleq \min\limits_{u_{c,k},w_{c,k}}\eta_{c,k}=g-R_{c,k}$ and $\eta_{k}^{\rm MMSE}\triangleq \min\limits_{u_{k},w_{k}}\eta_{k}=g-R_{k}$,
where $g=1/\ln2+\log_{2}\ln2$. Based on this, the problem $\textbf{{{P}}1}$ can be equivalently transformed into the WMMSE problem, which can be expressed as follows:
\begin{subequations}
	\begin{align}
	\textbf{{{P}}2-1:}  \min_{\substack{\boldsymbol{\rm F}_{\rm R F}, \boldsymbol{\rm F}_{\rm D},\boldsymbol{\rm c}, \\ \boldsymbol{\rm u}, \boldsymbol{\rm w}}}
	&\sum_{k \in \mathcal{K}} \alpha_{k}(\eta_{k}-C_{k}) \\
	\text { s.t. }
	& \sum_{i \in \mathcal{K}}C_{i}+\min_{u_{c,k},w_{c,k}}(\eta_{c,k})\leq g,\forall k \in \mathcal{K}, \label{C2-1}\\
	& \eqref{SCNR}, \eqref{Ck}-\eqref{units},
	\end{align}
\end{subequations}
where $\boldsymbol{\rm u}=\left[u_{c,1},\cdots,u_{c,K},u_{1},\cdots,u_{K}\right]^{T}$ denotes the vector of weights and $\boldsymbol{\rm w}=\left[w_{c,1},\cdots,w_{c,K},w_{1},\cdots,w_{K}\right]^{T}$ denotes the vector of equalizers. However, there is still the nonconvex constraint in (10c) of problem $\textbf{{{P}}2-1}$, which leads to the fact that $\textbf{{{P}}2-1}$ cannot be solved directly.

\subsection{Proposed PDD-Based Algorithm}
In the following, we propose the PDD-based algorithm to solve problem $\textbf{{{P}}2-1}$. We first introduce a set of auxiliary variables $\boldsymbol{\rm X}$, $\boldsymbol{\rm Y}$, $\{\boldsymbol{\rm Z}_{k}\}$, and $\{\boldsymbol{\rm q}_{k}\}$, $k \in \mathcal{K}$  that satisfy the following equality constraints: $\boldsymbol{\rm X}=\boldsymbol{\rm F}_{\rm RF}\boldsymbol{\rm F}_{\rm D}$, $\boldsymbol{\rm Y}=\boldsymbol{\rm X}$, $\boldsymbol{\rm Z}_{k}=\boldsymbol{\rm X}$, and $\boldsymbol{\rm q}_{k}=\boldsymbol{\rm c}$. And then, problem $\textbf{{{P}}2-1}$ can be equivalently converted to
\begin{subequations}
	\begin{align}
	\textbf{{{P}}2-2:}  \min_{\mathbb{X}}
	&\sum_{k \in \mathcal{K}} \alpha_{k}\eta_{k}(\boldsymbol{\rm X})-\boldsymbol{\rm \alpha}^{T}\boldsymbol{\rm c} \\
	\text { s.t. }
	&\frac{\operatorname{tr}\left(\boldsymbol{\rm Y}^{H} \boldsymbol{\rm \Phi} \boldsymbol{\rm Y}\right)} {\operatorname{tr}\left(\boldsymbol{\rm Y}^{H}\boldsymbol{\rm \Omega} \boldsymbol{\rm Y}\right)+N_{r}\sigma_{z}^{2}}\geq \gamma_{0}, \label{SCNR2-2}\\
	&\boldsymbol{\rm 1}^{T}\boldsymbol{\rm q}_{k}+\min_{u_{c,k},w_{c,k}}(\eta_{c,k}(\boldsymbol{\rm Z}_{k}))\leq g,\forall k \in \mathcal{K}, \label{C2-2}\\
	&\left\|\boldsymbol{\rm X}\right\|_{F}^{2} \leq P_{T},  \left|\boldsymbol{\rm F}_{\rm R F}(i, j)\right|=1,\forall i,j \label{PT2-2},\\
	&\boldsymbol{\rm c}\geq 0\label{cq2-2},\\
	&\boldsymbol{\rm X}=\boldsymbol{\rm F}_{\rm RF}\boldsymbol{\rm F}_{\rm D}, \boldsymbol{\rm Y}=\boldsymbol{\rm X}, \boldsymbol{\rm Z}_{k}=\boldsymbol{\rm X},\boldsymbol{\rm q}_{k}=\boldsymbol{\rm c},
	\end{align}
\end{subequations}
where $\mathbb{X}=\{\boldsymbol{\rm F}_{\rm R F}, \boldsymbol{\rm F}_{\rm D},\boldsymbol{\rm c},\boldsymbol{\rm u}, \boldsymbol{\rm w},\boldsymbol{\rm X}, \boldsymbol{\rm Y}, \boldsymbol{\rm Z}_{k}, \boldsymbol{\rm q}_{k}\}$ denotes a set of optimization variables and $\boldsymbol{\rm \alpha}=\left[\alpha_{1},\cdots,\alpha_{K} \right]^{T} $is a vector of user priority weight coefficients. 

According to the PDD algorithm, we introduce the Lagrange multipliers $\boldsymbol{\rm \Lambda}_{1}$, $\{\boldsymbol{\rm \Lambda}_{2,k}\}$, $\boldsymbol{\rm \Lambda}_{3}$, $\{\boldsymbol{\rm \lambda}_{c,k}\}$ and the penalty parameter $\rho$. Bringing the equality constraints (10f) to the OF, the resulting AL problem can be expressed as
\begin{subequations}
	\begin{align}
	\textbf{{{P}}2-3:}  \min_{\mathbb{X}}
	&\sum_{k \in \mathcal{K}} \alpha_{k}\eta_{k}(\boldsymbol{\rm X})-\boldsymbol{\rm \alpha}^{T}\boldsymbol{\rm c}+g_{\rho}(\mathbb{X})\\
	\text { s.t. }
	&\eqref{SCNR2-2}-\eqref{cq2-2},\label{2-3}
	\end{align}
\end{subequations}
where $g_{\rho}(\mathbb{X})=\frac{1}{2 \rho}(\|\boldsymbol{\rm X}-\boldsymbol{\rm Y}+\rho \boldsymbol{\rm \Lambda}_{1}\|_{F}^{2}+\sum_{k \in \mathcal{K}}\|\boldsymbol{\rm X}-\boldsymbol{\rm Z}_{k}+\rho \boldsymbol{\rm \Lambda}_{2,k}$
$\|_{F}^{2}+\|\boldsymbol{\rm X}-\boldsymbol{\rm F}_{\rm RF} \boldsymbol{\rm F}_{\rm D}+\rho \boldsymbol{\rm \Lambda}_{3}\|_{F}^{2}+\sum_{k \in \mathcal{K}}\left\|\boldsymbol{\rm c}-\boldsymbol{\rm q}_{k}+\rho \boldsymbol{\rm \lambda}_{c,k}\right\|_{F}^{2})$.
Clearly, when $\rho$ tends to 0, problem $\textbf{{{P}}2-3}$ is equivalent to problem $\textbf{{{P}}2-2}$. Next, fixing the Lagrange multipliers and the penalty parameter, we solve the inner loop problem using an iterative approach.
\subsection{Proposed Iterative Algorithm for Solving Problem \textbf{{{P}}2-3} }
According to the concave-convex procedure (CCCP) algorithm, the constraint (13b) can be transformed as
\begin{equation}
\begin{aligned}
f_{1}(\boldsymbol{\rm Y})-f_{2}(\boldsymbol{\rm Y})\leq 0,
\end{aligned}
\end{equation}
where $f_{1}(\boldsymbol{\rm Y})=\gamma_{0}\left(\operatorname{tr}\left(\boldsymbol{\rm Y}^{H}\boldsymbol{\rm \Omega} \boldsymbol{\rm Y}\right)+N_{r}\sigma_{z}^{2}\right)$ and $f_{2}(\boldsymbol{\rm Y})=\operatorname{tr}\left(\boldsymbol{\rm Y}^{H} \boldsymbol{\rm \Phi} \boldsymbol{\rm Y}\right)$.
At the $t$-th iteration, the first-order Taylor expansion of $f_{2}(\boldsymbol{\rm Y})$ at point $\boldsymbol{\rm Y}^{(t)}$, denoted by $\hat f_{2}(\boldsymbol{\rm Y},\boldsymbol{\rm Y}^{(t)})$, can be expressed as
\begin{equation}
\begin{aligned}
\hat f_{2}(\boldsymbol{\rm Y},\boldsymbol{\rm Y}^{(t)})\!=\!2 \mathfrak{R}\left\{\operatorname{tr}\left(\boldsymbol{\rm Y}^{(t) H} \boldsymbol{\rm \Phi} \boldsymbol{\rm Y}\right)\right\}\!\!-\!\!\operatorname{tr}\left(\boldsymbol{\rm Y}^{(t) H} \boldsymbol{\rm \Phi} \boldsymbol{\rm Y}^{(t)}\right).
\end{aligned}
\end{equation}
Thus, the constraint (13b) is approximately transformed into a convex constraint as follows:
\begin{equation}
\begin{aligned}
f_{1}(\boldsymbol{\rm Y})-\hat f_{2}(\boldsymbol{\rm Y},\boldsymbol{\rm Y}^{(t)})\leq 0.\label{SCNR_new}
\end{aligned}
\end{equation}
Based on the CCCP algorithm, problem $\textbf{{{P}}2-3} $ in the $t$-th iteration can be approximately given by
\begin{subequations}
	\begin{align}
	\textbf{{{P}}2-4:}  \min_{\mathbb{X}}
	&\sum_{k \in \mathcal{K}} \alpha_{k}\eta_{k}(\boldsymbol{\rm X})-\boldsymbol{\rm \alpha}^{T}\boldsymbol{\rm c}+g_{\rho}(\mathbb{X})\\
	\text { s.t. }
	&\eqref{C2-2}-\eqref{cq2-2},\eqref{SCNR_new}.
	\end{align}
\end{subequations}
In the following, we divide the optimized variables of problem $\textbf{{{P}}2-4}$ into five blocks, namely $\{\boldsymbol{\rm w}\}$, $\{\boldsymbol{\rm u}\}$, $\{\boldsymbol{\rm Y},\boldsymbol{\rm Z}_{k},\boldsymbol{\rm q}_{k},\boldsymbol{\rm F}_{\rm D}\}$, $\{\boldsymbol{\rm c},\boldsymbol{\rm F}_{\rm R F}\}$, and $\{\boldsymbol{\rm X}\}$. We define the accuracy tolerance $\epsilon^{in}$ and the maximum number of iterations $N_{max}^{in}$, initialize the iteration number $t=0$,  and repeat the steps until the difference between the OF values of two adjacent iterations is less than $\epsilon^{in}$ or $t\geq N_{max}^{in}$. The specific optimization steps for variable blocks are given as follows.

$\bm{Step 1}$: Optimize the equalizer vector $\{\boldsymbol{\rm w}\}$ by fixing other variables. Based on the derivation in Section~III.A, the optimal equalizers  can be updated as $w_{c,k}^{opt}=(\boldsymbol{\rm Z}_{k}\boldsymbol{\rm e}_{c})^{H} \boldsymbol{\rm h}_{k}(T_{c, k}(\boldsymbol{\rm Z}_{k}))^{-1}$ and $w_{k}^{opt}=(\boldsymbol{\rm X}\boldsymbol{\rm e}_{k})^{H} \boldsymbol{\rm h}_{k}(T_{k}(\boldsymbol{\rm X}))^{-1}$, where $T_{c, k}(\boldsymbol{\rm Z}_{k})$ and $T_{k}(\boldsymbol{\rm X})$ are obtained by taking $\boldsymbol{\rm F}_{\rm RF}\boldsymbol{\rm F}_{\rm D}=\boldsymbol{\rm X}=\boldsymbol{\rm Z}_{k}$ into $T_{c, k}$ and $T_{k}$, respectively.

$\bm{Step 2}$: Optimize the weight vector $\{\boldsymbol{\rm u}\}$ by fixing other variables. First, substitute $\boldsymbol{\rm F}_{\rm RF}\boldsymbol{\rm F}_{\rm D}=\boldsymbol{\rm X}=\boldsymbol{\rm Z}_{k}$ into Eqs. (9) to obtain $e_{c,k}^{\rm MMSE}(\boldsymbol{\rm Z}_{k})$ and $e_{k}^{\rm MMSE}(\boldsymbol{\rm X})$, respectively. Then, the optimal weights, denoted by $u_{c,k}^{opt}$ and $u_{k}^{opt}$, are given by $u_{c,k}^{opt}=(e_{c,k}^{\rm MMSE}(\boldsymbol{\rm Z}_{k})\ln 2)^{-1}$ and $u_{k}^{opt}=(e_{k}^{\rm MMSE}(\boldsymbol{\rm X})\ln 2)^{-1}$.

$\bm{Step 3}$: By fixing the other variables, optimize the block variables $\{\boldsymbol{\rm Y},\boldsymbol{\rm Z}_{k},\boldsymbol{\rm q}_{k},\boldsymbol{\rm F}_{\rm D}\}$. First, the subproblem for variable $\boldsymbol{\rm Y}$ can be expressed as
\begin{subequations}
	\begin{align}
    \min_{\boldsymbol{\rm Y}}  &\|\boldsymbol{\rm X}-\boldsymbol{\rm Y}+\rho \boldsymbol{\rm \Lambda}_{1}\|_{F}^{2}\\
	\text { s.t. }&\eqref{SCNR_new}.
	\end{align}
\end{subequations}
The optimal solution of problem~\eqref{SCNR_new}, denoted by $\boldsymbol{\rm Y}^{opt}(v_{1})$, can be obtained by one-dimensional line search, as follows:
\begin{align}
\boldsymbol{\rm Y}^{opt}(v_{1})\!=\!\left(\boldsymbol{\rm I}_{N_{t}}\!+\!v_{1} \gamma_{0} \boldsymbol{\rm \Omega}\right)^{-1}\left(\boldsymbol{\rm X}\!+\!\rho \boldsymbol{\rm \Lambda}_{1}\!+\!v_{1} \boldsymbol{\rm \Phi} \boldsymbol{\rm Y}^{(t)}\right), \label{eq:y_OUT}
\end{align}
where $v_{1}\geq 0$	is the Lagrange multiplier that needs to satisfy the KKT condition. Note that in the $(t+1)$ iteration, $\boldsymbol{\rm Y}^{(t)}$ is first updated by $\boldsymbol{\rm Y}^{(t)}=\boldsymbol{\rm Y}^{opt}$, and then $\boldsymbol{\rm Y}^{(t)}$ is taken into Eq.~\eqref{eq:y_OUT} to obtain the new $\boldsymbol{\rm Y}^{opt}$, i.e., updated in a recursive method.



Similarly, the optimal solutions of $\boldsymbol{\rm Z}_{k}$ and $\boldsymbol{\rm q}_{k}$, denoted by $\boldsymbol{\rm Z}_{k}^{opt}(v_{2,k})$ and $\boldsymbol{\rm q}_{k}^{opt}(v_{2,k})$, are given by
$\boldsymbol{\rm Z}_{k}^{opt}(v_{2,k})=\big(\boldsymbol{\rm I}_{N_{t}}+v_{2,k}u_{c, k}\left|w_{c, k}\right|^{2}\boldsymbol{\rm h}_{k}\boldsymbol{\rm h}_{k}^{H}\big)^{-1}\big(\boldsymbol{\rm X}+\rho \boldsymbol{\rm \Lambda}_{2,k}+v_{2,k}u_{c, k}w_{c,k}^{*}\boldsymbol{\rm h}_{k}\boldsymbol{\rm e}_{c}^{T}\big)$ and
$\boldsymbol{\rm q}_{k}^{opt}(v_{2,k})=\boldsymbol{\rm c}+\rho \boldsymbol{\rm \lambda}_{c,k}-\frac{v_{2,k}}{2} \boldsymbol{\rm 1}$,
where $v_{2,k}\geq 0$	denotes the Lagrange multiplier which can be obtained by the bisection method.

Next, the subproblem for variable $\boldsymbol{\rm F}_{\rm D}$ can be given by
\begin{align}
\min_{\boldsymbol{\rm F}_{\rm D}}  &\|\boldsymbol{\rm X}-\boldsymbol{\rm F}_{\rm RF} \boldsymbol{\rm F}_{\rm D}+\rho \boldsymbol{\rm \Lambda}_{3}\|_{F}^{2}.
\end{align}
By taking the first-order derivative of the variable $\boldsymbol{\rm F}_{\rm D}$ for the above subproblem (19), we obtain the optimal solution $\boldsymbol{\rm F}_{\rm D}^{opt}=\boldsymbol{\rm F}_{\rm RF}^{\dagger}\big(\boldsymbol{\rm X}+\rho \boldsymbol{\rm \Lambda}_{3}\big)$.

$\bm{Step 4}$: Optimize the block variables $\{\boldsymbol{\rm c},\boldsymbol{\rm F}_{\rm R F}\}$ by fixing other variables. The subproblem for the variable $\boldsymbol{\rm c}$ is given by
\begin{subequations}
	\begin{align}
	\min_{\boldsymbol{\rm c}}
	&\frac{1}{2 \rho}\sum_{k \in \mathcal{K}}\left\|\boldsymbol{\rm c}-\boldsymbol{\rm q}_{k}+\rho \boldsymbol{\rm \lambda}_{c,k}\right\|_{F}^{2}-\boldsymbol{\rm \alpha}^{T}\boldsymbol{\rm c}\\
	\text { s.t. }&\boldsymbol{\rm c}\geq 0.
	\end{align}
\end{subequations}
The problem (20) is a convex problem that can be easily optimized by the Karush-Kuhn-Tucker (KKT) condition. The optimal solution for the $i$-th element of the vector $\boldsymbol{\rm c}$, denoted by $C_{i}^{opt}, i \in \mathcal{K}$, has the closed form as follows:
\begin{align}
C_{i}^{opt}=\max \left\{0,\frac{\sum_{k \in \mathcal{K}}(\boldsymbol{\rm q}_{k,i}-\rho \boldsymbol{\rm \lambda}_{c,k,i})+\rho \alpha_{i}}{K}\right\},
\end{align}
where $\boldsymbol{\rm q}_{k,i}$ and $\boldsymbol{\rm \lambda}_{c,k,i}$ denote the $i$ th element of vectors $\boldsymbol{\rm q}_{k}$ and $\boldsymbol{\rm \lambda}_{c,k}$, respectively.

Next, the subproblem with respect to $\boldsymbol{\rm F}_{\rm RF}$ is given by
\begin{subequations}
\begin{align}
\min_{\boldsymbol{\rm F}_{\rm RF}}  &\|\boldsymbol{\rm X}-\boldsymbol{\rm F}_{\rm RF} \boldsymbol{\rm F}_{\rm D}+\rho \boldsymbol{\rm \Lambda}_{3}\|_{F}^{2}\\
\text { s.t. }& \left|\boldsymbol{\rm F}_{\rm R F}(i, j)\right|=1,\forall i,j.
\end{align}
\end{subequations}
To solve this non-convex quadratic programming problem, we transform it into a tractable form, as follows:
\begin{subequations}
\begin{align}
\min_{\boldsymbol{\rm F}_{\rm RF}}
&\operatorname{tr}\left(\boldsymbol{\rm F}_{\rm R F}^{H}\boldsymbol{\rm F}_{\rm R F}\boldsymbol{\rm C}\right)-2\mathfrak{R}\left\{\operatorname{tr}\left(\boldsymbol{\rm F}_{\rm R F}^{H}\boldsymbol{\rm B}\right)\right\}\\
\text { s.t. }& \left|\boldsymbol{\rm F}_{\rm R F}(i, j)\right|=1,\forall i,j,
\end{align}
\end{subequations}
where $\boldsymbol{\rm C}=\boldsymbol{\rm F}_{\rm D}\boldsymbol{\rm F}_{\rm D}^{H}$, $\boldsymbol{\rm B}=\left(\boldsymbol{\rm X}+\rho \boldsymbol{\rm \Lambda}_{3}\right)\boldsymbol{\rm F}_{\rm D}^{H}$. 


$\bm{Step 5}$: Optimize the variable $\boldsymbol{\rm X}$ by fixing other variables. Then the specific subproblem can be expressed as
\begin{subequations}
\begin{align}
\min_{\boldsymbol{\rm X}}&\sum_{k \in \mathcal{K}}\alpha_{k}u_{k}\left(\left|w_{k}\right|^{2}\sum_{i \in \mathcal{K}}\left|\boldsymbol{\rm h}_{k}^{H} \boldsymbol{\rm X}\boldsymbol{\rm e}_{i}\right|^{2}-2\mathfrak{R}\bigg\{w_{k} \boldsymbol{\rm h}_{k}^{H} \boldsymbol{\rm X}\boldsymbol{\rm e}_{k}\bigg\}\right)\notag\\
&+\frac{1}{2 \rho}\bigg(\|\boldsymbol{\rm X}-\boldsymbol{\rm Y}+\rho \boldsymbol{\rm \Lambda}_{1}\|_{F}^{2}+\sum_{k \in \mathcal{K}}\|\boldsymbol{\rm X}-\boldsymbol{\rm Z}_{k}+\rho \boldsymbol{\rm \Lambda}_{2,k}\|_{F}^{2}\notag\\
&+\|\boldsymbol{\rm X}-\boldsymbol{\rm F}_{\rm RF} \boldsymbol{\rm F}_{\rm D}+\rho \boldsymbol{\rm \Lambda}_{3}\|_{F}^{2}\bigg)\\
\text { s.t. }& \operatorname{tr}\left(\boldsymbol{\rm X}^{H}\boldsymbol{\rm X}\right)\leq P_{T}.
\end{align}
\end{subequations}
Problem (24) is a quadratic convex programming problem, and by introducing the Lagrange multiplier $v_{3}\geq 0$, we can obtain the optimal $\boldsymbol{\rm X}^{opt}(v_{3})$, which is given by
$\boldsymbol{\rm x}^{opt}(v_{3})\!=\!\operatorname{vec}\left(\boldsymbol{\rm X}^{o p t}(v_{3})\right)\!=\!\big((\frac{K+2}{2 \rho}\!+\!v_{3}) \boldsymbol{\rm I}_{N_{t} \times(K+1)}\!+\!(\sum_{i \in \mathcal{K}} \boldsymbol{\rm e}_{i} \boldsymbol{\rm e}_{i}^{H})^{T} \!\otimes\!(\sum_{k \in \mathcal{K}} \alpha_{k} u_{k}\left|w_{k}\right|^{2} \boldsymbol{\rm h}_{k} \boldsymbol{\rm h}_{k}^{H})
\big)^{\!-\!1} \operatorname{vec}(\boldsymbol{\rm V})$, where
$\boldsymbol{\rm V}=\sum_{k \in \mathcal{K}} \alpha_{k} u_{k} w_{k}^{*} \boldsymbol{\rm h}_{k} \boldsymbol{\rm e}_{k}^{H}+\frac{1}{2 \rho}\big(\boldsymbol{\rm Y}+\sum_{k \in \mathcal{K}} \boldsymbol{\rm Z}_{k}+\boldsymbol{\rm F}_{\rm RF} \boldsymbol{\rm F}_{\rm D}-\rho(\boldsymbol{\rm \Lambda}_{1}+\sum_{k \in \mathcal{K}} \boldsymbol{\rm \Lambda}_{2,k}+\boldsymbol{\rm \Lambda}_{3})\big)$, and $\otimes$ denotes the Kronecker products between two matrices.
The Lagrange multiplier $v_{3}$ can be obtained by the bisection method.



\subsection{Summary of the Proposed WMMSE-PDD Algorithm}
After each inner iteration is completed, we need to update the Lagrange multipliers and the penalty parameter of the outer loop. We first define the constraint violation in the $m$-th outer iteration, denoted by $h\left(\mathbb{X}^{(m)}\right)$, expressed by
$h\left(\mathbb{X}^{(m)}\right)\!=\!\max\{\|\boldsymbol{\rm X}\!-\!\boldsymbol{\rm Y}\|_{F},\|\boldsymbol{\rm X}\!-\!\boldsymbol{\rm Z}_{k}\|_{F},\|\boldsymbol{\rm c}\!-\!\boldsymbol{\rm q}_{k}\|_{2}, \|\boldsymbol{\rm X}\!-\!\boldsymbol{\rm F}_{\rm RF} \boldsymbol{\rm F}_{\rm D}\|_{F}\}$.
In the $(m+1)$-th outer iteration, we introduce the variable $\zeta^{(m+1)}$ associated with the constraint violation $h(\mathbb{X}^{(m)})$, usually set to $\zeta^{(m+1)}=0.9h(\mathbb{X}^{(m)})$, and then give the update rule as follows: when $h(\mathbb{X}^{(m+1)})\geq\zeta^{(m+1)}$, the penalty parameter can updated according to $\rho^{(m+1)}=c_{\rho}\rho^{(m)}(0<c_{\rho}<1)$; otherwise, the Lagrange multipliers need to be updated, which can be given by
\begin{equation}
\begin{cases}
&\hspace{-0.4cm}\boldsymbol{\rm \Lambda}_{1}^{(m+1)}\!=\!\boldsymbol{\rm \Lambda}_{1}^{(m)}\!+\!\frac{\boldsymbol{\rm X}^{(m)}-\boldsymbol{\rm Y}^{(m)}}{\rho^{(m)}}, \boldsymbol{\rm \Lambda}_{2,k}^{(m+1)}\!=\!\boldsymbol{\rm \Lambda}_{2,k}^{(m)}\!+\!\frac{\boldsymbol{\rm X}^{(m)}\!-\!\boldsymbol{\rm Z}_{k}^{(m)}}{\rho^{(m)}};\\
&\hspace{-0.4cm}\boldsymbol{\rm \Lambda}_{3}^{(m+1)}\!\!=\!\!\boldsymbol{\rm \Lambda}_{3}^{(m)}\!+\!\frac{\boldsymbol{\rm X}^{(m)}\!-\!\boldsymbol{\rm F}_{\rm RF}^{(m)} \boldsymbol{\rm F}_{\rm D}^{(m)}}{\rho^{(m)}}, \boldsymbol{\rm \lambda}_{c,k}^{(m+1)}\!=\!\boldsymbol{\rm \lambda}_{c,k}^{(m)}\!\!+\!\!\frac{\boldsymbol{\rm c}^{(m)}\!-\!\boldsymbol{\rm q}_{k}^{(m)}}{\rho^{(m)}}.
\end{cases}
\end{equation}
In summary, the detailed steps of the proposed WMMSE-PDD algorithm for handling joint common rate allocation and hybrid beamforming design are listed in $\textbf{Algorithm 1}$. The complexity of our proposed WMMSE-PDD algorithm is approximately $\mathcal{O}\big(N_{max}^{out}N_{max}^{in}\big((N_{t}(K+1))^{3}+I_{\rm RF}N_{t}^{2}(K+1)^{2}+K\log_{2}(\frac{I_{0}}{\epsilon})\big)\big)$,
where $N_{max}^{in}$ and$N_{max}^{out}$  denote the number of inner and outer loops, respectively.

\begin{algorithm}[h]
	\caption{Proposed WMMSE-PDD Algorithm for Joint Common Rate Allocation and HBF Design} \label{alg2}
	\begin{algorithmic}[1]
		\STATE Define the accuracy tolerance $\epsilon^{out}$ and the maximum number of iterations $N_{max}^{out}$. Initialize $\mathbb{X}^{(0)}$, $\boldsymbol{\rm \Lambda}^{(0)}$, $\rho^{(0)}$, $\zeta^{0}$, set $0<c_{\rho}<1$, the iteraion number $m=0$.
		\REPEAT
		\STATE Update $\mathbb{X}^{(m+1)}$ according to Section~III-C;
		\IF{$h(\mathbb{X}^{(m+1)})\leq\zeta^{(m)}$}
		\STATE Update $\boldsymbol{\rm \Lambda}^{(m+1)}$ by applying Eq.(25),\\ $\rho^{(m+1)}=\rho^{(m)}$;
		\ELSE
		\STATE $\boldsymbol{\rm \Lambda}^{(m+1)}=\boldsymbol{\rm \Lambda}^{(m)}$, $\rho^{(m+1)}=c_{\rho}\rho^{(m)}$;
		\ENDIF
		\STATE $\zeta^{(m+1)}=0.9h(\mathbb{X}^{(m)})$
		\STATE $m\gets m+1$;
		\UNTIL $h(\mathbb{X}^{(m)})\leq\epsilon^{out}$ or $m\geq   N_{max}^{out}$.
	\end{algorithmic}
\end{algorithm}

\vspace{-0.2cm}
\section{Numerical Results}
In this section, we provide numerical results to demonstrate the performance of the proposed RSMA-enhanced mmWave ISAC system with HBF structure. Throughout the whole simulation, we set the carrier frequency as 28 $\mathrm{GHz}$, $N_{t}=32$, $N_{r}=32$, and $K=4$.  Unless  mentioned otherwise, we consider the user noise $\sigma_{{k}}^{2}=-100\ \mathrm{dBm}$ and the sensing echo noise $\sigma_{{z}}^{2}=-100\ \mathrm{dBm}$. The weight factor is set as $\alpha_{k}=1$, and the input SCNR threshold for target sensing is set as $\overline{\gamma}_{r}^{in}=10\ \mathrm{dB}$. The system power budget is $P_{T}=30\ \mathrm{dBm}$. Further, to compare the performance of the system under different mmWave channel conditions: low spatial correlation channel and high spatial correlation channel.
\begin{figure}[htbp]
	\centering
	\includegraphics[scale=0.45]{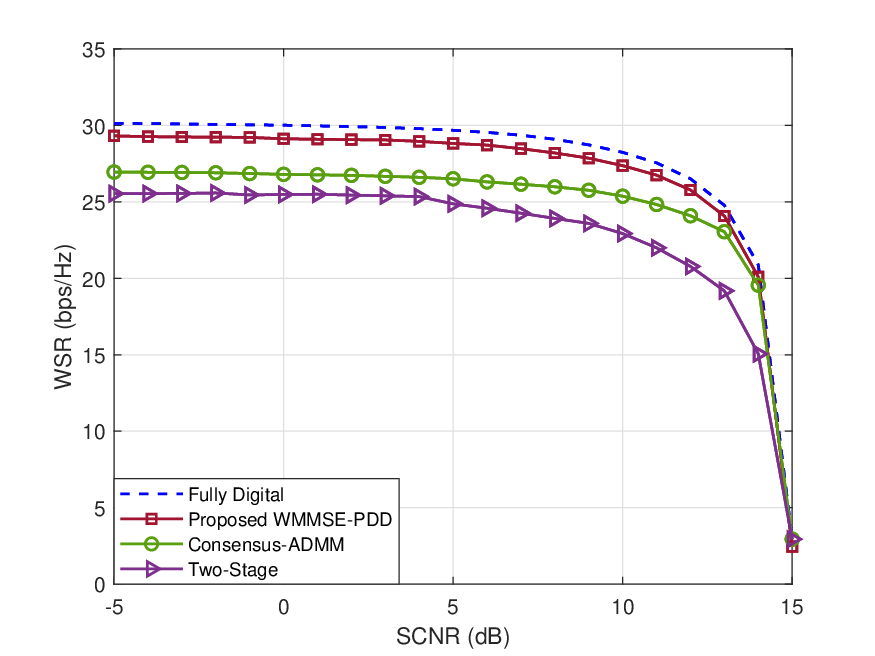}
	\vspace{-0.3cm}
	\caption{WSR by different methods versus sensing SCNR.} \label{fig:the algorithm}
	\vspace{-0.3cm}
\end{figure}


To verify the superiority of the proposed WMMSE-PDD algorithm, we compare it in Fig. 2 with the following benchmark algorithms: 1) the Consensus-ADMM algorithm \cite{ChengZiyang2}\cite{sensors2024}, which is a distributed optimization method; and 2) the ``Two-Stage" algorithm\cite{YuXianghao}, which seeks to minimize the Euclidean distance of the optimal fully-digital beamforming matrix. It can be shown that, the proposed algorithm outperforms the above two benchmark algorithms in terms of the communication-sensing achievable performance region.

\begin{figure}[htbp]
	\centering
	\includegraphics[scale=0.45]{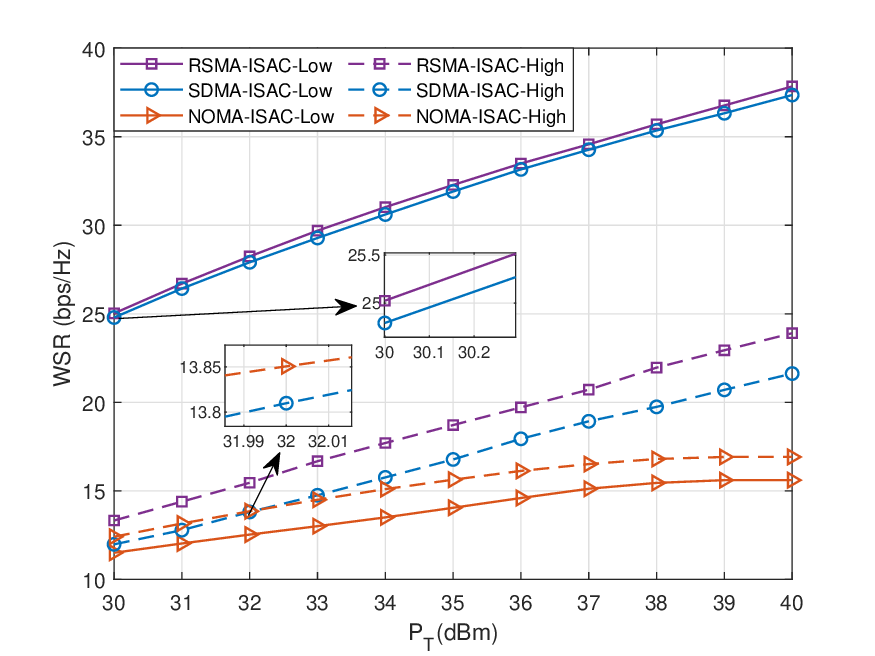}
	\vspace{-0.35cm}
	\caption{WSR $R_{\rm WSR}$ versus power budget $P_{T}$, where $\gamma_{0}=10\ {\rm  dB}$.} \label{fig: WSR VS PT}
	\includegraphics[scale=0.45]{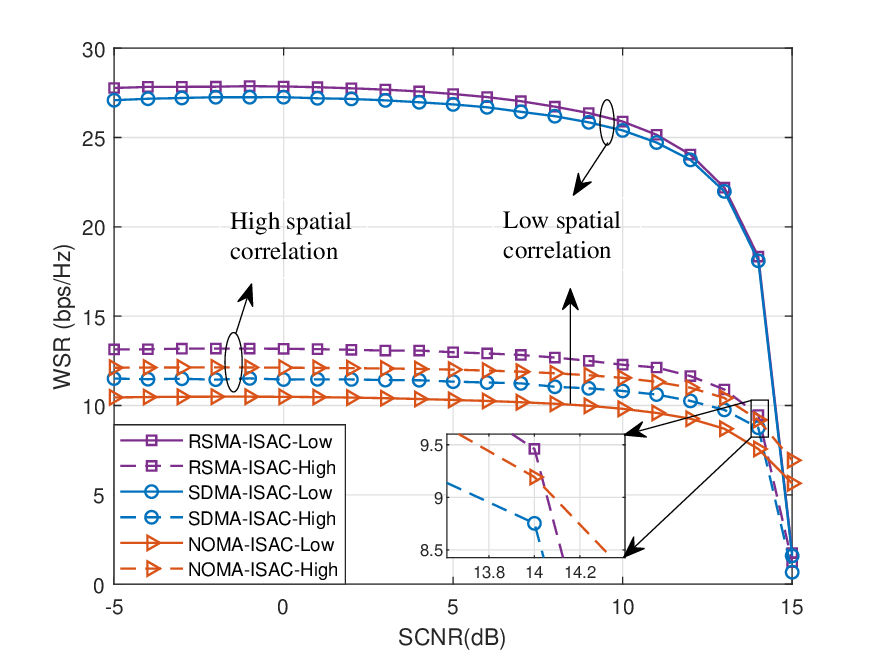}
	\vspace{-0.35cm}
	\caption{WSR $R_{\rm WSR}$ versus sensing SCNR $\gamma_{0}$, where $P_{T}=30\ {\rm dBm}$.} \label{fig: WSR VS SCNR}
\vspace{-0.3cm}
\end{figure}

Figures 3 and 4 show user WSR versus budget power  and sensing SCNR threshold, respectively. For both high and low spatial correlation channels, the RSMA-ISAC scheme outperforms the SDMA-ISAC and NOMA-ISAC schemes in terms of WSR and anti-clutter interference capability. We reveal that the common data stream has a dual function in mitigating multi-user interference and trade-offs between communication and sensing performance. For low spatial correlation channels, it can be seen that the WSR of the RSMA-ISAC scheme is slightly higher than that of the SDMA-ISAC scheme but significantly higher than that of the NOMA-ISAC scheme. The reason is that under the high spatial DoF of mmWave channels configured with massive antennas, the strong user can fully decode the weak user message in the NOMA scheme, thus leading to the loss of multiplexing gain and rate. In summary, the common data stream of our proposed RSMA-ISAC scheme can reduce interference between users and further improve WSR. Note that when the sensing performance is extremely demanding, i.e., $\gamma_{0}= 15\ {\rm dB}$ in the Fig.4, the WSR of the NOMA-ISAC scheme outperforms both RSMA-ISAC and NOMA-ISAC, which is attributed to the fact that the SIC decoding strategy is used for each user in the NOMA scheme, which also provides a certain performance gain in the case of scarce communication resources.

\section{Conclusions}\label{sec:Conclusion}
In this paper, we have investigated the problem of joint common rate allocation and HBF design for the RSMA-assisted mmWave ISAC system. We have proposed the WMMSE-PDD algorithm to solve this nonlinear and nonconvex optimization problem with coupled constraints. Numerical results demonstrated that our proposed HBF-based RSMA-assisted mmWave ISAC system has the advantages of higher energy efficiency, more robust interference management, and better communication-sensing performance trade-offs.

\bibliographystyle{IEEEtran}
\bibliography{References}

\end{document}